\documentclass[aps,prb,showpacs,twocolumn,superscriptaddress,floatfix]{revtex4}

\usepackage{graphicx}
\usepackage{dcolumn}
\usepackage{bm}
\usepackage{amsmath}
\usepackage{amssymb}

\begin{document}

\renewcommand{\vec}[1]{\mbox{\boldmath $#1$}}


\title{Modified Cross-Correlation for Efficient White-Beam Inelastic Neutron Scattering Spectroscopy}


\author{K. Tomiyasu}
\email[Electronic address: ]{tomiyasu@m.tohoku.ac.jp}
\affiliation{Department of Physics, Tohoku University, Aoba, Sendai 980-8578, Japan}
\author{M. Matsuura}
\affiliation{Institute for Materials Research, Tohoku University, Aoba, Sendai 980-8577, Japan}
\author{H. Kimura}
\affiliation{Institute for Multidisciplinary Research for Advanced Materials, Tohoku University, Aoba, Sendai 980-8577, Japan}
\author{K. Iwasa}
\affiliation{Department of Physics, Tohoku University, Aoba, Sendai 980-8578, Japan}
\author{K. Ohoyama}
\affiliation{Institute for Materials Research, Tohoku University, Aoba, Sendai 980-8577, Japan}
\author{T. Yokoo}
\affiliation{Neutron Science Laboratory, High Energy Accelerator Organization, Oho, Tsukuba 305-0801, Japan}
\author{S. Itoh}
\affiliation{Neutron Science Laboratory, High Energy Accelerator Organization, Oho, Tsukuba 305-0801, Japan}
\author{E. Kudoh}
\affiliation{Department of Information and Communication Engineering, Tohoku Institute of Technology, Taihaku, Sendai 982-8577, Japan}
\author{T. Sato}
\affiliation{WPI Advanced Institute for Materials Research, Tohoku University, Aoba, Sendai 980-8577, Japan}
\author{M. Fujita}
\affiliation{Institute for Materials Research, Tohoku University, Aoba, Sendai 980-8577, Japan}



\date{\today}

\begin{abstract}
We describe a method of white-beam inelastic neutron scattering for improved measurement efficiency. The method consists of matrix inversion and selective extraction. The former is to resolve each incident energy component from the white-beam data, and the latter eliminates contamination by elastic components, which produce strong backgrounds that otherwise obfuscate the inelastic scattering components. In this method, the optimal experimental condition to obtain high efficiency will strongly depend on the specific aim of the individual experiments.
\end{abstract}


\maketitle

\section{Introduction}
Inelastic neutron scattering has come to be recognized as indispensable in modern materials science, because a material's spin and lattice dynamics provides unique information about a system's Hamiltonian. A complete description of these excitations in momentum ($\vec{Q}$) and energy ($E$) space is needed to fully reconstruct the interactions that govern a material's behavior on the atomic scale. However, the technique normally requires a large volume of sample, often on the order of several cubic centimeters~\cite{Tomiyasu_2009}. This is a very significant limitation in research to develop new materials with novel functions.

\begin{figure}[htbp]
\begin{center}
\includegraphics[width=3.2in, keepaspectratio]{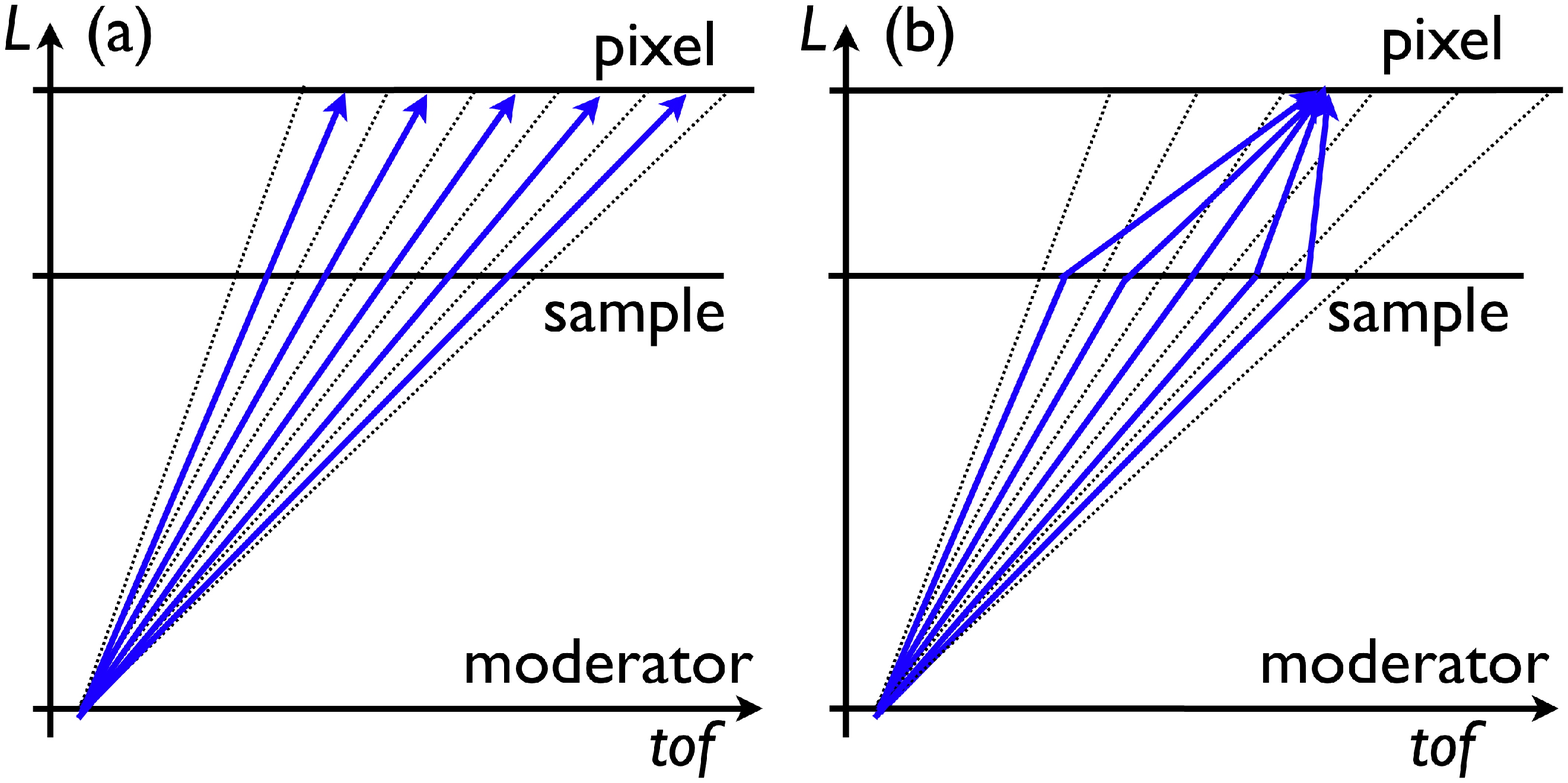}
\end{center}
\caption{\label{fig:dgm_old} (Color online) Plot of TOF against position without sequence chopper. Solid arrows indicate the most probable neutrons. Dotted lines sectionalize each $E_{\rm i}$ channel. (a) Diffraction. (b) Inelastic scattering. }
\end{figure}

In neutron diffraction experiments, however, the development of time-of-flight (TOF) technique allowed the use of a white beam for increase in measurement efficiency compared with the conventional method using a monochromatic beam. Each wavelength (energy) component can be resolved by TOF, as shown in Fig.~\ref{fig:dgm_old}(a), and finally merged into a single diffraction $|\vec{Q}|$ pattern for a powder sample (time focusing) or a $\vec{Q}$ map for a single-crystal sample.
Unfortunately, the same cannot be applied to inelastic scattering, because the different $E_{\rm i}$ (incident energy) and $E_{\rm f}$ (final energy) components are entangled at the same TOF for the same pixel on the detector, as shown in Fig.~\ref{fig:dgm_old}(b). Thus, it has been considered that either $E_{\rm i}$ or $E_{\rm f}$ must be monochromatized or must be analyzed, either of which incurs a large loss in neutron intensity.

\begin{figure}[htbp]
\begin{center}
\includegraphics[width=3.2in, keepaspectratio]{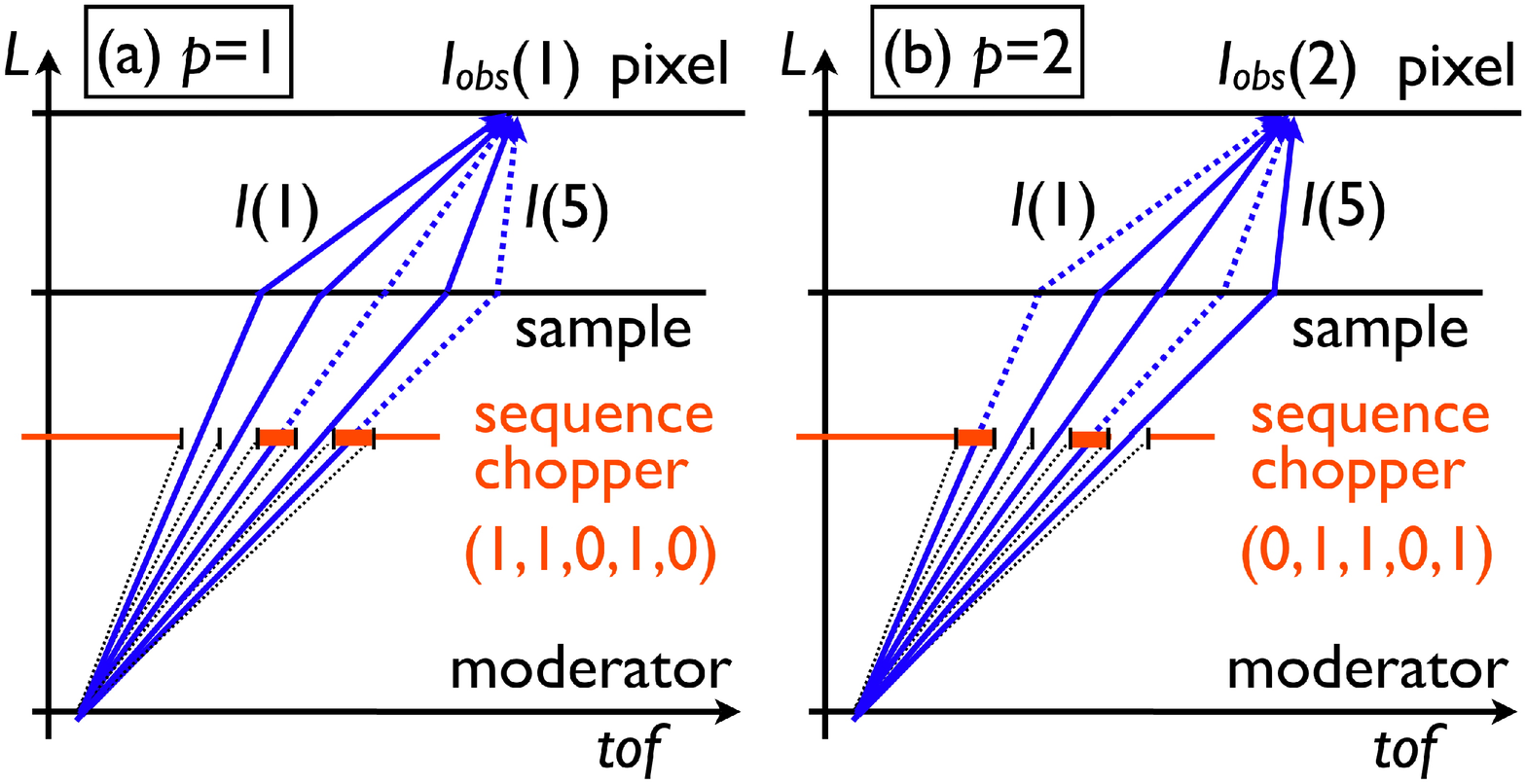}
\end{center}
\caption{\label{fig:dgm_XC} (Color online) Example of TOF--position diagrams with sequence chopper ($N=5$) at sequence phases $p=1$ (a) and $p=2$ (b). By cyclically shifting the phase, a raw data set of $I_{\rm obs}(1)$, $I_{\rm obs}(2)$, $I_{\rm obs}(3)$, $I_{\rm obs}(4)$, and $I_{\rm obs}(5)$ is obtained at a specific TOF and at a specific pixel on the detector. After the measurements, each $E_{\rm i}$ component of $I(1)$, $I(2)$, $I(3)$, $I(4)$, and $I(5)$ can be resolved mathematically. }
\end{figure}

Another remarkable method, called cross-correlation, was developed over four decades ago as an extension to the white-beam diffraction~\cite{Skold_1968}. The method basically involves extracting the elastic components and removing the inelastic components~\cite{Pellionisz_1971}. As shown in Fig.~\ref{fig:dgm_XC}, a special mechanical chopper modulates a white incident pulsed beam with a pseudorandom open/close sequence, and $N$-times cyclic phase shifts of the modulation generate a set of $N$ data with different $E_{\rm i}$ contrast. Then, on the basis of the contrast, the data for each $E_{\rm i}$ can be mathematically resolved.

The mathematical formalization is given below. Here, for convenience, parameters and functions are renamed and redefined from those in the original papers~\cite{Skold_1968,Pellionisz_1971}. The intensity detected at a specific TOF at a specific pixel of the detector, $I_{\rm obs}(p)$ $(p=1,\dots,N)$, is described by
\begin{equation}
\label{eq:corr_1}
I_{\rm obs}(p) = \sum_{j=1}^{N}F(j+p)I(j) + B
\end{equation}
where $p$ is the phase shift of the sequence (phase of sequence chopper), $F(k)$ is the $k$-th element in the sequence $F$ consisting of only 0 (close) and 1 (open), $F(k+N)$ is defined to be equal to $F(k)$ for $k=1, \dots, N$, $j$ is an index for $E_{\rm i}$, $I(j)$ is the intensity coming from the $j$-th $E_{\rm i}$ component in a white incident pulsed beam for $j=1, \dots, N$, and $B$ is the background. The pseudorandom sequence $F$ is restricted by
\begin{subequations}
\begin{eqnarray}
\label{eq:corr_2a}
N = 2^{n} -1 \phantom{4} (n: \rm{integer}), \\
\label{eq:corr_2b}
F^{\prime}(k) = 2F(k)-1, \\
\label{eq:corr_2c}
\sum_{k=1}^{N}F^{\prime}(k) = 1, \\
\label{eq:corr_2d}
\sum_{k=1}^{N}F^{\prime}(k)F^{\prime}(k+k^{\prime}) = (N+1)\delta_{0,k^{\prime}} -1.
\end{eqnarray}
\end{subequations}
%
This type of sequence $F$ is currently called a maximum length sequence, which is generated by a simple recurrence formula and is widely applied in the field of digital communications~\cite{Golomb_1967}. Combining the above equations, one can resolve each $E_{\rm i}$ component,
\begin{equation}
I(j) = \frac{2}{N+1}\sum_{p=1}^{N}F^{\prime}(j+p)I_{\rm obs}(p) - \frac{2}{N+1}B.
\label{eq:corr_3}
\end{equation}

It should be noted, however, that the method cannot be directly applied to inelastic scattering because the elastic components and their large statistical errors obfuscate the very weak inelastic components. This is probably why the method has not been realized thus far in an actual instrument dedicated to inelastic scattering. In fact, for the new CORELLI instrument under construction at the Spallation Neutron Source (SNS) in Oak Ridge, Tennessee, the method will be used mainly to study diffusive elastic scattering such as in frustrated systems and ionic conductors~\cite{Rosenkranz_2008}.

This paper presents a modification to this method for a more practical white-beam inelastic neutron scattering setup. The proposed method has two novel aspects: the introduction of an inverse matrix representation and a proposed method for selective extraction. The former affords a different solution to Eq.~(\ref{eq:corr_1}) to resolve each $E_{\rm i}$ component in the white-beam data. The latter eliminates contamination by elastic components, which otherwise produce strong backgrounds. Finally, we present estimates of some instrumental specifications for the TOF polarized neutron spectrometer, POLANO, being constructed at J-PARC.

\section{Inverse matrix representation}
We formalize an alternative solution to Eq.~(\ref{eq:corr_1}) without the conditions Eqs.~(\ref{eq:corr_2a})--(\ref{eq:corr_2d}). Here, the measurement of $I_{\rm obs}(p)$ $(p=1, \dots, N)$ is the same as in the original method except for the kind of sequence. Ignoring $B$ for simplicity, Eq.~(\ref{eq:corr_1}) can be represented by
\begin{equation}
\vec{I}_{\rm obs} = \hat{F} \vec{I},
\label{eq:tomi_1}
\end{equation}
where $\vec{I}_{\rm obs}$ is the vector $\left[{I}_{\rm obs}(p)\right]$ $(p=1, \dots, N)$; $\hat{F}$ is the matrix $\left[\vec{F}_{p=1}, \vec{F}_{p=2}, \dots, \vec{F}_{p=N}\right]$; $\vec{F}_p$ is the sequence vector $\left[ F(j+p) \right]$ $(j=1, \dots, N)$; and $\vec{I}$ is the vector $\left[I(j)\right]$. Hence, one can resolve $\vec{I}$ by
\begin{equation}
\vec{I} = \hat{F}^{-1} \vec{I}_{\rm obs}.
\label{eq:tomi_2}
\end{equation}

Consider, for example, the sequence (0, 1, 1, 0, 1) at $p=1$:
\begin{eqnarray}
\left[
\begin{array}{c}
I_{\rm obs}(1) \\
I_{\rm obs}(2) \\
I_{\rm obs}(3) \\
I_{\rm obs}(4) \\
I_{\rm obs}(5) \\
\end{array}
\right]
=
\left[
\begin{array}{ccccc}
0 & 1 & 1 & 0 & 1 \\
1 & 0 & 1 & 1 & 0 \\
0 & 1 & 0 & 1 & 1 \\
1 & 0 & 1 & 0 & 1 \\
1 & 1 & 0 & 1 & 0 \\
\end{array}
\right]
\left[
\begin{array}{c}
I(1) \\
I(2) \\
I(3) \\
I(4) \\
I(5) \\
\end{array}
\right].
\label{eq:tomi_3}
\end{eqnarray}
Hence, one can obtain
\begin{eqnarray}
\left[
\begin{array}{c}
I(1) \\
I(2) \\
I(3) \\
I(4) \\
I(5) \\
\end{array}
\right]
=
\frac{1}{3}
\left[
\begin{array}{ccccc}
-1 & -1 & -1 & 2 & 2 \\
2 & -1 & -1 & -1 & 2 \\
2 & 2 & -1 & -1 & -1 \\
-1 & 2 & 2 & -1 & -1 \\
-1 & -1 & 2 & 2 & -1 \\
\end{array}
\right]
\left[
\begin{array}{c}
I_{\rm obs}(1) \\
I_{\rm obs}(2) \\
I_{\rm obs}(3) \\
I_{\rm obs}(4) \\
I_{\rm obs}(5) \\
\end{array}
\right].
\label{eq:tomi_4}
\end{eqnarray}

Thus, almost all types of sequences can be used as long as $F^{-1}$ exists. Taking into account $B$ again, one can also identify a sequence to minimize $|F^{-1}(B,B,B,B,B)|$, for example, by trial and error with many numerical trials.

It should be noted that this general matrix formalization is not considered superior to the maximum length sequence. However, the general matrix formalization does afford an advantage when the conditions of Eqs.~(\ref{eq:corr_2a})$-$(\ref{eq:corr_2d}) are not satisfied on actual instrumentation, for example, because of insufficient switching speed between 0 and 1 for high $E_{\rm i}$ range or high resolution. In this paper, we use the general matrix formalization only because selective extraction, proposed in the next section, also does not fulfill the conditions.

\section{Selective extraction}
We explain the proposed selective extraction method using the above example in Eq.~(\ref{eq:tomi_3}). First, one needs to prepare another chopper with the inverted sequence--from open/close to close/open, that is, from 1/0 to 0/1. The inverted chopper gives another data set,
\begin{eqnarray}
\left[
\begin{array}{c}
J_{\rm obs}(1) \\
J_{\rm obs}(2) \\
J_{\rm obs}(3) \\
J_{\rm obs}(4) \\
J_{\rm obs}(5) \\
\end{array}
\right]
=
\left[
\begin{array}{ccccc}
1 & 0 & 0 & 1 & 0 \\
0 & 1 & 0 & 0 & 1 \\
1 & 0 & 1 & 0 & 0 \\
0 & 1 & 0 & 1 & 0 \\
0 & 0 & 1 & 0 & 1 \\
\end{array}
\right]
\left[
\begin{array}{c}
I(1) \\
I(2) \\
I(3) \\
I(4) \\
I(5) \\
\end{array}
\right],
\label{eq:tomi_5}
\end{eqnarray}
where $\left[ J_{\rm obs}(p) \right]$ is the raw data obtained for the phase $p$. Then, we consider the case where $I(3)$ is the elastic component for the targeted TOF and pixel of the detector. Our purpose is to remove $I(3)$. Thus, by selectively extracting only the arrays in which the third column is 0 (sequence chopper closed) from Eqs.~(\ref{eq:tomi_3}) and (\ref{eq:tomi_5}), one can reconstruct a good quality data set:
\begin{eqnarray}
\left[
\begin{array}{c}
J_{\rm obs}(1) \\
J_{\rm obs}(2) \\
I_{\rm obs}(3) \\
J_{\rm obs}(4) \\
I_{\rm obs}(5) \\
\end{array}
\right]
=
\left[
\begin{array}{ccccc}
1 & 0 & 0 & 1 & 0 \\
0 & 1 & 0 & 0 & 1 \\
0 & 1 & 0 & 1 & 1 \\
0 & 1 & 0 & 1 & 0 \\
1 & 1 & 0 & 1 & 0 \\
\end{array}
\right]
\left[
\begin{array}{c}
I(1) \\
I(2) \\
I(3) \\
I(4) \\
I(5) \\
\end{array}
\right].
\label{eq:tomi_6}
\end{eqnarray}
Hence,
\begin{eqnarray}
\left[
\begin{array}{c}
J_{\rm obs}(1) \\
J_{\rm obs}(2) \\
I_{\rm obs}(3) \\
J_{\rm obs}(4) \\
I_{\rm obs}(5) \\
\end{array}
\right]
=
\left[
\begin{array}{cccc}
1 & 0 & 1 & 0 \\
0 & 1 & 0 & 1 \\
0 & 1 & 1 & 1 \\
0 & 1 & 1 & 0 \\
1 & 1 & 1 & 0 \\
\end{array}
\right]
\left[
\begin{array}{c}
I(1) \\
I(2) \\
I(4) \\
I(5) \\
\end{array}
\right].
\label{eq:tomi_7}
\end{eqnarray}
This equation can be solved by dropping one array and using the inverse matrix, or by the least squares method.

Further, we can reincorporate the background term $B$:
\begin{eqnarray}
\left[
\begin{array}{c}
J_{\rm obs}(1) \\
J_{\rm obs}(2) \\
I_{\rm obs}(3) \\
J_{\rm obs}(4) \\
I_{\rm obs}(5) \\
\end{array}
\right]
=
\left[
\begin{array}{ccccc}
1 & 0 & 1 & 0 & 1 \\
0 & 1 & 0 & 1 & 1 \\
0 & 1 & 1 & 1 & 1 \\
0 & 1 & 1 & 0 & 1 \\
1 & 1 & 1 & 0 & 1 \\
\end{array}
\right]
\left[
\begin{array}{c}
I(1) \\
I(2) \\
I(4) \\
I(5) \\
B \\
\end{array}
\right].
\label{eq:tomi_8}
\end{eqnarray}
Hence,
\begin{eqnarray}
\left[
\begin{array}{c}
I(1) \\
I(2) \\
I(4) \\
I(5) \\
B \\
\end{array}
\right]
=
\left[
\begin{array}{ccccc}
0 & 0 & 0 & -1 & 1 \\
-1 & 0 & 0 & 0 & 1 \\
0 & -1 & 1 & 0 & 0 \\
0 & 0 & 1 & -1 & 0 \\
1 & 1 & -1 & 1 & -1 \\
\end{array}
\right]
\left[
\begin{array}{c}
J_{\rm obs}(1) \\
J_{\rm obs}(2) \\
I_{\rm obs}(3) \\
J_{\rm obs}(4) \\
I_{\rm obs}(5) \\
\end{array}
\right].
\label{eq:tomi_9}
\end{eqnarray}

We emphasize that the selective extraction method is applicable not only when the third channel is elastic but also when an arbitrary channel is elastic. One can remove the elastic components at all TOFs and pixels with only the two data sets.

\section{Statistical efficiency}
Price and Sk$\ddot{\rm o}$ld reported that the cross-correlation method is not always better than the conventional monochromatic method in terms of statistical efficiency~\cite{Price_1970}. The reason is that white-beam data counts $\left[ I_{\rm obs}(p) \right]$ and $\left[ J_{\rm obs}(p) \right]$ are inevitably large and are accompanied by large statistical errors $\left[ \Delta I_{\rm obs}(p) \right]=\left[ \sqrt{I_{\rm obs}(p)} \right]$ and $\left[ \Delta J_{\rm obs}(p) \right]=\left[ \sqrt{J_{\rm obs}(p)} \right]$, which are propagated to $\left[ I(j) \right]$ with further enhancement by adding and subtracting $\left[ I_{\rm obs}(p) \right]$ and $\left[ J_{\rm obs}(p) \right]$. For example, $\{\Delta I(1)\}^{2} = (-1/3)^{2} \cdot I_{\rm obs}(1) + (-1/3)^{2} \cdot I_{\rm obs}(2) + (-1/3)^{2} \cdot I_{\rm obs}(3) + (2/3)^{2} \cdot I_{\rm obs}(4) + (2/3)^{2} \cdot I_{\rm obs}(5)$ in Eq.~(\ref{eq:tomi_4}), and $\{\Delta I(1)\}^{2} = (-1)^{2} \cdot J_{\rm obs}(4) + 1^{2} \cdot I_{\rm obs}(5)$ in Eq.~(\ref{eq:tomi_9}), where $\Delta I(1)$ denotes a statistical error of $I(1)$. 
In addition, since $N$ measurements with cyclic rotation of sequence are needed to resolve $\left[ I_{\rm obs}(p) \right]$ to $\left[ I(j) \right]$ with only one-measurement statistics, the use of white beam does not overall multiply the measurement efficiency so much. 
In the use of maximum length sequence, a statistical advantage can be obtained only for special $j$(s) ($E_{\rm i}$ channel(s)), in which the signal of interest is more than twice the average counts per channel~\cite{Price_1970}. Hence, the cross-correlation methods would be suitable only in the cases of phonon resonance, magnon resonance, or elastic scattering. 

The above situation is essentially the same as that in our modified cross-correlation method with selective extraction. Here, we roughly estimate the statistical efficiency based on the assumption that $\left[ I(j) \right]$ consists of only inelastic scattering counts with similar magnitude and without huge elastic components. 
First, in the cross-correlation methods without selective extraction including the original one and the aforementioned general-matrix one, the error of $I^{(\rm nsl)}(j)$ is described by
\begin{equation}
\begin{split}
\{\Delta I^{(\rm nsl)}(j)\}^{2} = \sum_{p=1}^{N} (\hat{A}^{-1}_{jp})^{2} I_{\rm obs}^{(\rm nsl)}(p) \\
\sim a_{j}^{2} \langle I_{\rm obs}^{(\rm nsl)} \rangle
\sim N_{\rm open} a_{j}^{2} \langle I^{(\rm nsl)} \rangle,
\end{split}
\label{eq:err_1a}
\end{equation}
where the superscript (\rm nsl) denotes the non-use of selective extraction, $\hat{A}^{-1}$ corresponds to the inverse matrix, $a_{j}^{2} = \sum_{p=1}^{N}(\hat{A}^{-1}_{jp})^{2}$,
$\langle I_{\rm obs}^{(\rm nsl)} \rangle = \sum_{p=1}^{N}I_{\rm obs}^{(\rm nsl)}(p)/N$,
$N_{\rm open}$ is the number of opening channels ($\sim N/2$),
and $\langle I^{(\rm nsl)} \rangle = \sum_{j=1}^{N}I^{(\rm nsl)}(j)/N$.
Hence, the statistical efficiency $\eta_{j}^{(\rm nsl)}$ is estimated by
\begin{equation}
\begin{split}
\eta_{j}^{(\rm nsl)} \equiv \frac{I^{(\rm nsl)}(j)}{\Delta I^{(\rm nsl)}(j)} \\
\sim \left( \frac{c_{j}}{\sqrt{N_{\rm open}}a_{j}} \right) \cdot \sqrt{\langle I^{(\rm nsl)} \rangle} \\
\sim \left( \frac{\sqrt{2}c_{j}}{\sqrt{N}a_{j}} \right) \cdot \sqrt{\langle I^{(\rm nsl)} \rangle},
\end{split}
\label{eq:err_2a}
\end{equation}
where $I^{(\rm nsl)}(j) \equiv c_{j}\langle I^{(\rm nsl)} \rangle$. 
%
%
Next, in the modified cross-correlation method, the error of $I^{(\rm sl)}(j)$ is described by
\begin{equation}
\begin{split}
\{\Delta I^{(\rm sl)}(j)\}^{2} = \sum_{p=1}^{N} (\hat{B}^{-1}_{j^{\prime}p})^{2} \{K_{\rm obs}^{(\rm sl)}(p)\} \\
\sim \sum_{p=1}^{N} (\hat{B}^{-1}_{j^{\prime}p})^{2} \{I_{\rm obs}^{(\rm nsl)}(p)/2\} \\
\sim \frac{1}{2} b_{j^{\prime}}^{2} \langle I_{\rm obs}^{(\rm nsl)} \rangle 
\sim \frac{1}{2} N_{\rm open} b_{j^{\prime}}^{2} \langle I^{(\rm nsl)} \rangle,
\end{split}
\label{eq:err_1b}
\end{equation}
where the superscript (sl) denotes the use of selective extraction, $K_{\rm obs}^{(\rm sl)}(p)=I_{\rm obs}^{(\rm sl)}(p)$ or $J_{\rm obs}^{(\rm sl)}(p)$ like in Eq.~(\ref{eq:tomi_9}), $I_{\rm obs}^{(\rm sl)}(p) \sim J_{\rm obs}^{(\rm sl)}(p) \sim I_{\rm obs}^{(\rm nsl)}(p)/2$ since selective extraction involves the use of two data sets: one from the original chopper and the other from the inverted chopper, $\hat{B}^{-1}$ corresponds to the final inverse matrix like in Eq.~(\ref{eq:tomi_9}), $b_{j^{\prime}}^{2} = \sum_{p=1}^{N}(\hat{B}^{-1}_{j^{\prime}p})^{2}$, and $j^{\prime}$ denotes the row number in $\hat{B}^{-1}$ with which $I^{(\rm sl)}(j)$ can be obtained (e.g., $j^{\prime}=1$ for $j=1$ and $j^{\prime}=3$ for $j=4$ in Eq.~(\ref{eq:tomi_9})). 
Hence, the statistical efficiency $\eta_{j}^{(\rm sl)}$ is estimated by
\begin{equation}
\begin{split}
\eta_{j}^{(\rm sl)} \equiv \frac{I^{(\rm sl)}(j)}{\Delta I^{(\rm sl)}(j)} 
= \frac{I^{(\rm nsl)}(j)/2}{\Delta I^{(\rm sl)}(j)} \\
\sim \left( \frac{c_{j}}{\sqrt{2}\sqrt{N_{\rm open}}b_{j^{\prime}}} \right) \cdot \sqrt{\langle I^{(\rm nsl)} \rangle} \\
\sim \left( \frac{c_{j}}{\sqrt{N}b_{j^{\prime}}} \right) \cdot \sqrt{\langle I^{(\rm nsl)} \rangle},
\end{split}
\label{eq:err_2b}
\end{equation}
where $I^{(\rm sl)}(j)=I^{(\rm nsl)}(j)/2$ since selective extraction involves the use of two data sets again. 
Then, in the conventional monochromatic experiments, the efficiency $\eta_{j}^{\rm (mono)}$ required to obtain the same data set of all $E_{\rm i}$ ($j=1, \dots, N$) over the same total measurement time using monochromatic beam is estimated by
\begin{equation}
\eta_{j}^{\rm (mono)} 
= \frac{c_{j}\langle I^{(\rm nsl)} \rangle}{\sqrt{c_{j}\langle I^{(\rm nsl)} \rangle}} 
= \sqrt{c_{j}} \cdot \sqrt{\langle I^{(\rm nsl)} \rangle}, 
\label{eq:err_3}
\end{equation}
which corresponds to the measurements described by $\hat{A}=\hat{E}$ (identity matrix) in the general matrix formalization. 
Thus, the ratios of the statistical efficiencies are estimated by
\begin{subequations}
\begin{eqnarray}
\frac{ \eta_{j}^{(\rm nsl)} }{ \eta_{j}^{\rm (mono)} } \sim \sqrt{ \frac { 2c_{j} } { Na_{j}^2 } }, 
\label{eq:err_4a} \\
\frac{ \eta_{j}^{(\rm sl)} }{ \eta_{j}^{\rm (mono)} } \sim \sqrt{ \frac { c_{j} } { Nb_{j^{\prime}}^2 } }.
\label{eq:err_4b}
\end{eqnarray}
\end{subequations}

Using example sequences summarized in the Appendix, we numerically calculated the values of $a_{j}$ and $b_{j^{\prime}}$ and the criteria of $c_{j}$ to retrieve the $I(j)$ of interest  more efficiently than the use of conventional monochromatic beam. 
For the original cross-correlation method with maximum length sequence without selective extraction, the results are $a_{j}^{2}=0.44$, 0.23, 0.12, 0.062, 0.031, and 0.016 for all $j$ at $N=7$, 15, 31, 63, 127, and 255, respectively. Hence, $(\eta_{j}^{(\rm nsl)} / \eta_{j}^{\rm (mono)})>1$ gives the criteria of $c_{j}>1.5$, 1.8, 1.9, 2.0, 2.0, and 2.0, which are consistent with Price's criterion of $c_{j}>2$ for all the $N$; the $I(j)$ of interest should be more than twice the average counts per channel~\cite{Price_1970}. 
For the modified cross-correlation method, $b_{j^{\prime}}^{2}=0.88$, 0.39, 0.22, 0.15, 0.096, and 0.059 ($j^{\prime} \neq N$: $I(j)$) and $b_{N}^{2}=0.28$, 0.31, 0.63, 0.78, 0.20, and 0.30 ($j^{\prime} = N$: constant background). Hence, $(\eta_{j}^{(\rm sl)} / \eta_{j}^{\rm (mono)})>1$ ($j \neq N$: $I(j)$) gives $c_{j}>6.2$, 5.8, 6.7, 9.4, 12, and 15. The latter criteria are harder than the former ones in the absence of huge elastic components, as expected. In addition, since $c_{j}$ increases as $N$ increases in the latter, it would be better to set $N$ less than about 60. 

In this way, both the original and modified cross-correlation methods can efficiently give only the $I(j)$ components with relatively large $c_{j}$ among $\left[ I(j) \right]$. Therefore, it is important to tune the experimental conditions, such as the ranges of $E_{\rm i}$, TOF, and pixel used, such that the components of interest become the strongest in intensity among $\left[ I(j) \right]$. In this sense, we would like to emphasize that the modified method with selective extraction can remove not only an elastic channel but also an arbitrary one channel of no interest with relatively strong intensity among $\left[ I(j) \right]$; for example, spurious neutrons scattered on unexpected paths and spurious neutrons coming from the previous flame. 

In practice, (1) for elastic scattering, the original method without selective extraction could be safely applied, as has been reported in the past~\cite{Skold_1968, Pellionisz_1971, Rosenkranz_2008, Price_1970}, since elastic scattering is normally the strongest among $\left[ I(j) \right]$. However, the modified method might improve the data as an insurance to remove a spurious channel. 
(2) For quasielastic scattering and low-energy inelastic scattering, which will have the next strongest intensity and will necessarily overlap with the strongest elastic scattering in TOF, the modified method will be effective to remove the elastic scattering. 
(3) For higher energy modes with relatively weak intensity, one must remove not only an elastic channel but also quasielastic and low-energy inelastic channels, which contaminate the higher energy data as a source of statistical errors. Therefore, it would be better to narrow the $E_{i}$ range from white to quasi-monochromatic, which will allow us to avoid all the elastic, quasielastic, and low energy inelastic scatterings by TOF, in conjunction with the modified method to remove an arbitrary channel of no interest again. 
In any case, to efficiently utilize the cross-correlation methods, users should recognize that the optimal experimental condition will highly depend on the specific aim of the individual experiments.

\section{Specifications for practical implementation}
We are constructing a TOF polarized neutron spectrometer called POLANO at a decoupled moderator at J-PARC. Because polarization devices such as Heusler crystals and spin filters lose a large proportion of neutrons, we need a method to gain an increase of over a factor of 10 in measurement efficiency. Our method of white-beam inelastic neutron scattering is one candidate, though it is applicable to both unpolarized and polarized inelastic neutron scattering. In this section, taking this spectrometer as an example, we present some specifications for the sequence chopper.

For this system, we set the distance between moderator and sample $L_{1}$ as 17.0 m, the distance between sample and detector $L_{2}$ as 2.0 m, and the distance between sequence chopper and sample $L_{3}$ as 2.0 m. The time width values at the decoupled moderator at J-PARC, $\Delta t_{\rm m}$, were used for each $E_{\rm i}$ (Table~\ref{tab:dt_sc}). For each $E_{\rm i}$, the energy resolution $\Delta E/E_{i}$ was kept $\leq0.04$. Then, using the analytical formula for energy resolution~\cite{Tomiyasu_2006}, we evaluated the opening time per $E_{\rm i}$ channel required at the sequence chopper, $\Delta t_{\rm sc1}$.

\begin{figure}[htbp]
\begin{center}
\includegraphics[width=2.1in, keepaspectratio]{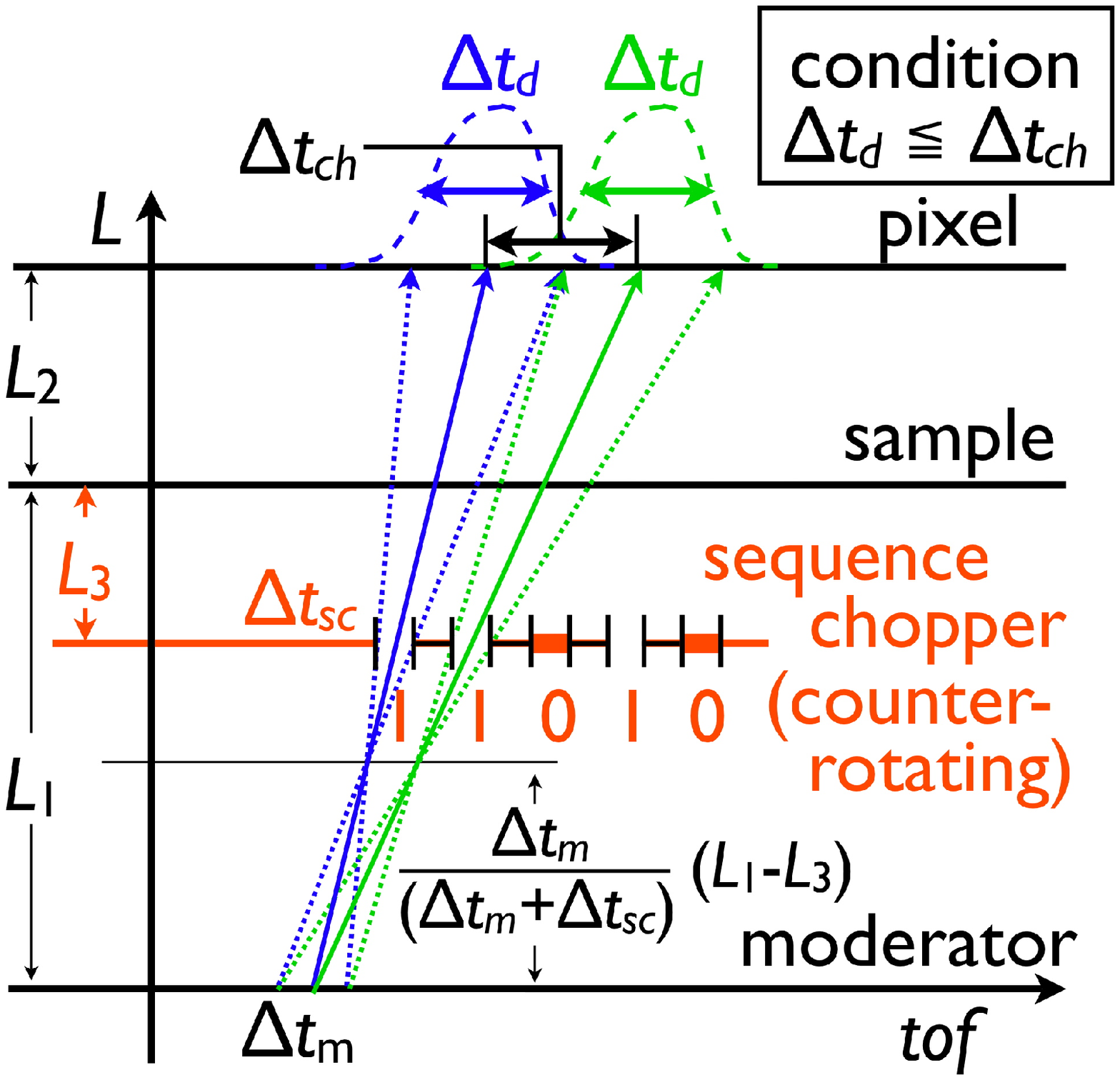}
\end{center}
\caption{\label{fig:dgm_new} (Color online) Diagram showing TOF and position with sequence chopper in the estimated specifications for POLANO. Solid arrows indicate the most probable neutrons, and dotted arrows indicate most inaccurate neutrons. The sequence chopper is alternately closed using a set of counter-rotating disk choppers. The alternating closing parts do not generate the sequence. }
\end{figure}

Also, on the assumption that a sequence chopper is alternately closed without generating a sequence as an example, as shown in Fig.~\ref{fig:dgm_new}, we estimated the condition of $\Delta t_{\rm sc2}$ necessary to avoid contamination by the elastic tails from neighboring $E_{\rm i}$ channels. The condition can be estimated by $\Delta t_{\rm d} \leq \Delta t_{\rm ch}$, where $\Delta t_{\rm d}$ is the time width of the elastic component at the pixel on the detector with $E_{\rm i}$, and $\Delta t_{\rm ch}$ is the time difference between two TOFs of neighboring $E_{\rm i}$ channels at the pixel, as defined in Fig.~\ref{fig:dgm_new}. The two parameters are described by other parameters,
\begin{subequations}
\begin{eqnarray}
\Delta t_{\rm d} &=& \frac{ L_{1}+L_{2} - \frac{\Delta t_{\rm m}}{\Delta t_{\rm m}+\Delta t_{\rm sc2}}(L_{1}-L_{3}) }{\frac{\Delta t_{\rm m}}{\Delta t_{\rm m}+\Delta t_{\rm sc2}}(L_{1}-L_{3})} \cdot \Delta t_{\rm m}, \\
\Delta t_{\rm ch} &=& \frac{L_{1}+L_{2}}{L_{1}-L_{3}} \cdot (2\Delta t_{\rm sc2}) \phantom{2}.
\end{eqnarray}
\end{subequations}

As summarized in Table~\ref{tab:dt_sc}, there exists a solution for $\Delta t_{\rm sc}$. For example, over a wide $E_{\rm i}$ range of 10--80 meV, $\Delta t_{\rm sc} = 9$ $\mu$sec simultaneously satisfies the constraint of $\Delta E/E_{i} \leq 0.04$ and the avoidance of the tails from neighboring $E_{\rm i}$ channels.
This opening time can be realized by a set of counter-rotating disk choppers with the following parameters: 700 mm-$\phi$, 20 mm/channel, and 350 Hz~\cite{comm_1}.
In addition, for the sequence chopper, note that an arbitrary sequence can be generated by printing a sequence clockwise and then counterclockwise on counter-rotating disk choppers.  The sequence chopper generates $0.5\cdot(\pi\cdot700 \phantom{1} \rm{(mm)})/20 \phantom{1} \rm{(mm)}=55$ $E_{\rm i}$ channels.
\begin{table}[htbp]
\begin{center}
\caption{\label{tab:dt_sc} Results of numerical estimation of system specifications for POLANO. The values $\Delta t_{\rm sc1}$ and $\Delta t_{\rm sc2}$ are obtained so as to satisfy $\Delta E/E_{i} \leq 0.04$ and avoid contamination by the elastic tails from neighboring $E_{\rm i}$ channels, respectively. All the time widths are defined as full widths at half maximum. }
\bigskip
\begin{tabular}{cccc}
\hline
$E_{\rm i}$ (meV) & $\Delta t_{\rm m}$ ($\mu$sec) & $\Delta t_{\rm sc1}$ ($\mu$sec)& $\Delta t_{\rm sc2}$ ($\mu$sec)\\
\hline
5.0 & 50 & $\leq31$ & $\geq11$ \\
10 & 32 & $\leq22$ & $\geq7$ \\
20 & 20 & $\leq16$ & $\geq4$ \\
40 & 13 & $\leq11$ & $\geq3$ \\
80 & 9 & $\leq9$ & $\geq2$ \\
\hline
\end{tabular}
\end{center}
\end{table}
\section{Summary}
We developed a modified cross-correlation method for an increase in measurement efficiency of inelastic neutron scattering. First, a different solution using an inverse matrix representation was formalized to resolve each $E_{\rm i}$ component in the white-beam data. Second, a method of selective extraction was proposed to avoid contamination by elastic components. Third, taking spectrometer POLANO at J-PARC as an example, practical specifications for the sequence chopper were estimated. 
Experimental situations giving high efficiency, however, would be quite limited and complex. 
\appendix*
The example sequences used for the calculations in Sec. 4 are summarized. For the original cross-correlation method, 
(1, 0, 0, 1, 1, 1, 0) at $N=7$, 
(1, 0, 1, 1, 0, 0, 1, 0, 0, 0, 1, 1, 1, 1, 0) at $N=15$, 
(0, 1, 1, 0, 1, 0, 0, 1, 0, 0, 0, 0, 1, 0, 1, 0, 1, 1, 1, 0, 1, 1, 0, 0, 0, 1, 1, 1, 1, 1, 0) at $N=31$, 
(1, 0, 1, 0, 1, 1, 0, 0, 1, 1, 0, 1, 1, 1, 0, 1, 1, 0, 1, 0, 0, 1, 0, 0, 1, 1, 1, 0, 0, 0, 1, 0, 1, 1, 1, 1, 0, 0, 1, 0, 1, 0, 0, 0, 1, 1, 0, 0, 0, 0, 1, 0, 0, 0, 0, 0, 1, 1, 1, 1, 1, 1, 0) at $N=63$, 
(1, 0, 1, 0, 1, 0, 0, 1, 1, 0, 0, 1, 1, 1, 0, 1, 1, 1, 0, 1, 0, 0, 1, 0, 1, 1, 0, 0, 0, 1, 1, 0, 1, 1, 1, 1, 0, 1, 1, 0, 1, 0, 1, 1, 0, 1, 1, 0, 0, 1, 0, 0, 1, 0, 0, 0, 1, 1, 1, 0, 0, 0, 0, 1, 0, 1, 1, 1, 1, 1, 0, 0, 1, 0, 1, 0, 1, 1, 1, 0, 0, 1, 1, 0, 1, 0, 0, 0, 1, 0, 0, 1, 1, 1, 1, 0, 0, 0, 1, 0, 1, 0, 0, 0, 0, 1, 1, 0, 0, 0, 0, 0, 1, 0, 0, 0, 0, 0, 0, 1, 1, 1, 1, 1, 1, 1, 0) at $N=127$, 
(1, 1, 0, 1, 1, 0, 0, 1, 1, 1, 1, 0, 0, 0, 1, 1, 0, 1, 0, 1, 1, 1, 0, 0, 1, 0, 0, 0, 0, 1, 1, 1, 1, 0, 1, 1, 1, 0, 1, 1, 1, 1, 0, 1, 0, 0, 0, 0, 0, 1, 0, 0, 0, 0, 0, 0, 1, 0, 1, 1, 0, 1, 1, 1, 1, 1, 0, 0, 1, 1, 0, 1, 1, 1, 0, 0, 0, 1, 0, 1, 1, 1, 0, 1, 0, 0, 1, 1, 0, 0, 1, 0, 1, 0, 1, 0, 1, 0, 0, 1, 0, 0, 1, 0, 0, 0, 1, 0, 1, 0, 0, 0, 0, 1, 0, 0, 1, 1, 0, 1, 0, 0, 0, 1, 1, 1, 1, 1, 0, 1, 0, 1, 1, 0, 1, 0, 0, 1, 0, 1, 0, 0, 1, 1, 1, 1, 1, 1, 0, 0, 0, 0, 0, 0, 0, 1, 1, 0, 1, 1, 0, 1, 0, 1, 0, 0, 0, 1, 0, 0, 1, 0, 1, 1, 1, 1, 0, 0, 1, 0, 1, 1, 0, 0, 0, 1, 0, 0, 0, 1, 1, 0, 0, 1, 1, 0, 0, 0, 1, 1, 1, 0, 0, 0, 0, 1, 1, 0, 0, 0, 0, 0, 1, 1, 1, 0, 1, 1, 0, 0, 0, 0, 1, 0, 1, 0, 1, 1, 0, 0, 1, 0, 0, 1, 1, 1, 0, 0, 1, 1, 1, 0, 1, 0, 1, 0, 1, 1, 1, 1, 1, 1, 1, 1, 0) at $N=255$. 

For the modified cross-correlation method, the original chopper sequences are 
(1, 1, 1, 1, 0, 1, 1) at $N=7$, 
(1, 0, 0, 1, 1, 0, 1, 0, 1, 1, 1, 1, 1, 1, 1) at $N=15$, 
(1, 0, 1, 0, 0, 1, 0, 0, 0, 1, 0, 1, 0, 1, 1, 0, 1, 0, 0, 0, 0, 0, 0, 1, 1, 0, 0, 0, 1, 0, 1) at $N=31$, 
(1, 1, 0, 0, 0, 0, 1, 1, 0, 0, 0, 1, 0, 0, 0, 1, 1, 0, 1, 1, 0, 1, 1, 0, 1, 1, 0, 1, 1, 0, 0, 0, 0, 1, 1, 0, 0, 0, 0, 0, 1, 1, 0, 0, 0, 0, 0, 0, 0, 1, 1, 1, 0, 1, 0, 0, 0, 0, 0, 1, 1, 1, 1) at $N=63$, 
(0, 1, 1, 0, 1, 0, 0, 1, 0, 0, 0, 0, 1, 1, 1, 1, 0, 0, 1, 0, 1, 1, 0, 0, 0, 1, 1, 1, 1, 0, 1, 0, 0, 0, 0, 0, 1, 0, 1, 0, 0, 0, 1, 0, 0, 0, 1, 1, 0, 1, 0, 0, 1, 0, 0, 1, 1, 0, 0, 1, 1, 1, 0, 1, 0, 1, 0, 0, 0, 1, 0, 0, 1, 0, 0, 0, 0, 0, 0, 1, 0, 0, 0, 1, 1, 0, 0, 1, 0, 1, 1, 1, 1, 0, 1, 1, 1, 0, 0, 0, 0, 0, 0, 0, 1, 1, 0, 0, 0, 1, 0, 0, 0, 1, 1, 0, 1, 0, 1, 0, 0, 0, 0, 1, 0, 0, 0) at at $N=127$, 
(0, 0, 1, 1, 0, 0, 1, 0, 1, 0, 0, 1, 1, 0, 1, 0, 0, 0, 1, 0, 0, 1, 1, 1, 0, 0, 0, 1, 0, 1, 1, 1, 1, 0, 0, 1, 0, 0, 1, 0, 1, 0, 1, 1, 1, 1, 0, 0, 0, 0, 0, 1, 1, 1, 0, 0, 0, 1, 0, 1, 1, 1, 0, 0, 0, 0, 0, 0, 0, 0, 0, 1, 0, 0, 1, 1, 1, 1, 0, 0, 1, 0, 0, 1, 0, 0, 1, 1, 0, 0, 0, 1, 0, 1, 1, 1, 0, 1, 1, 0, 0, 1, 1, 1, 0, 0, 0, 1, 0, 0, 0, 0, 1, 1, 1, 0, 1, 1, 0, 0, 1, 1, 0, 0, 0, 1, 0, 1, 0, 0, 0, 1, 0, 0, 0, 1, 0, 1, 0, 0, 0, 1, 0, 0, 1, 1, 0, 0, 0, 0, 0, 1, 1, 0, 0, 1, 1, 1, 0, 1, 1, 1, 1, 0, 0, 1, 0, 0, 0, 0, 0, 0, 1, 1, 0, 1, 1, 0, 1, 0, 0, 0, 1, 0, 1, 0, 0, 1, 0, 0, 1, 1, 0, 0, 0, 0, 1, 1, 0, 1, 0, 0, 0, 1, 0, 1, 0, 0, 1, 1, 1, 1, 0, 0, 1, 1, 0, 0, 0, 1, 1, 0, 0, 0, 0, 0, 1, 1, 0, 1, 1, 0, 1, 1, 0, 0, 1, 1, 1, 0, 0, 1, 0, 1, 1, 0, 0, 0, 0, 0, 1, 1, 1, 0, 1) at $N=255$.   

%
%
\acknowledgments
We thank Dr S. Rosenkranz for the helpful discussion and Dr H. Hiraka, Professor Y. Noda, and Professor K. Yamada for their encouragement. This study was financially supported by Grants-in-Aid for Young Scientists (B) (22740209), Priority Areas (22014001), and Scientific Researches (S) (21224008) and (A) (22244039) from the MEXT of Japan.
The study was also supported by the Neutron Scattering Program Advisory Committee of IMSS, KEK (2009S09).

\bibliography{WINS_6_arXiv}

\begin{thebibliography}{8}
\expandafter\ifx\csname natexlab\endcsname\relax\def\natexlab#1{#1}\fi
\expandafter\ifx\csname bibnamefont\endcsname\relax
  \def\bibnamefont#1{#1}\fi
\expandafter\ifx\csname bibfnamefont\endcsname\relax
  \def\bibfnamefont#1{#1}\fi
\expandafter\ifx\csname citenamefont\endcsname\relax
  \def\citenamefont#1{#1}\fi
\expandafter\ifx\csname url\endcsname\relax
  \def\url#1{\texttt{#1}}\fi
\expandafter\ifx\csname urlprefix\endcsname\relax\def\urlprefix{URL }\fi
\providecommand{\bibinfo}[2]{#2}
\providecommand{\eprint}[2][]{\url{#2}}

\bibitem[{\citenamefont{Tomiyasu et~al.}(2009)\citenamefont{Tomiyasu, Fujita,
  Kolesnikov, Bewley, Bull, and Bennington}}]{Tomiyasu_2009}
\bibinfo{author}{\bibfnamefont{K.}~\bibnamefont{Tomiyasu}},
  \bibinfo{author}{\bibfnamefont{M.}~\bibnamefont{Fujita}},
  \bibinfo{author}{\bibfnamefont{A.~I.} \bibnamefont{Kolesnikov}},
  \bibinfo{author}{\bibfnamefont{R.~I.} \bibnamefont{Bewley}},
  \bibinfo{author}{\bibfnamefont{M.~J.} \bibnamefont{Bull}}, \bibnamefont{and}
  \bibinfo{author}{\bibfnamefont{S.~M.} \bibnamefont{Bennington}},
  \bibinfo{journal}{Appl. Phys. Lett.} \textbf{\bibinfo{volume}{94}},
  \bibinfo{pages}{092502} (\bibinfo{year}{2009}), \bibinfo{note}{and references
  therein}.

\bibitem[{\citenamefont{Sk$\ddot{\rm{o}}$ld}(1968)}]{Skold_1968}
\bibinfo{author}{\bibfnamefont{K.}~\bibnamefont{Sk$\ddot{\rm{o}}$ld}},
  \bibinfo{journal}{Nucl. Instrum. Methods} \textbf{\bibinfo{volume}{63}},
  \bibinfo{pages}{114} (\bibinfo{year}{1968}).

\bibitem[{\citenamefont{\text{E. g., }P. Pellionisz}(1971)}]{Pellionisz_1971}
\bibinfo{author}{\bibnamefont{\text{E. g., }P. Pellionisz}},
  \bibinfo{journal}{Nucl. Instrum. Methods} \textbf{\bibinfo{volume}{92}},
  \bibinfo{pages}{125} (\bibinfo{year}{1971}).

\bibitem[{\citenamefont{Golomb}(Holden-Day, Inc., San Francisco,
  1967)}]{Golomb_1967}
\bibinfo{author}{\bibfnamefont{S.~W.} \bibnamefont{Golomb}},
  \bibinfo{journal}{{\it Shift Register Sequences}}
  (\bibinfo{year}{Holden-Day, Inc., San Francisco, 1967}).

\bibitem[{\citenamefont{Rosenkranz and Osborn}(2008)}]{Rosenkranz_2008}
\bibinfo{author}{\bibfnamefont{S.}~\bibnamefont{Rosenkranz}} \bibnamefont{and}
  \bibinfo{author}{\bibfnamefont{R.}~\bibnamefont{Osborn}},
  \bibinfo{journal}{Pramana J. Phys.} \textbf{\bibinfo{volume}{71}},
  \bibinfo{pages}{705} (\bibinfo{year}{2008}).

\bibitem[{\citenamefont{Price and Sk$\ddot{\rm o}$ld}(1970)}]{Price_1970}
\bibinfo{author}{\bibfnamefont{D.~L.} \bibnamefont{Price}} \bibnamefont{and}
  \bibinfo{author}{\bibfnamefont{K.}~\bibnamefont{Sk$\ddot{\rm o}$ld}},
  \bibinfo{journal}{Nucl. Inst. and Methods} \textbf{\bibinfo{volume}{82}},
  \bibinfo{pages}{208} (\bibinfo{year}{1970}).

\bibitem[{\citenamefont{Tomiyasu and Itoh}(2006)}]{Tomiyasu_2006}
\bibinfo{author}{\bibfnamefont{K.}~\bibnamefont{Tomiyasu}} \bibnamefont{and}
  \bibinfo{author}{\bibfnamefont{S.}~\bibnamefont{Itoh}},
  \bibinfo{journal}{Physica B} \textbf{\bibinfo{volume}{385-386}},
  \bibinfo{pages}{1110} (\bibinfo{year}{2006}), \bibinfo{note}{and references
  therein}.

\bibitem[{com()}]{comm_1}
\bibinfo{note}{The value of 9 $\mu$sec is obtained by $\sqrt{8\ln{2}/12} \cdot
  20 \rm{(mm/channel)}/\{ (\pi \cdot 700 \rm{(mm/cycle)}) (2 \cdot 350
  \rm{(cycles/sec)}) \}$.}

\end{thebibliography}

\end{document}